\documentclass[12pt]{article}
\usepackage{amsmath}
\usepackage{graphicx}
\usepackage{natbib}
\usepackage{url} 
\usepackage{makecell} 

\usepackage{authblk} 

\newcommand{\blind}{0}

\usepackage[utf8]{inputenc}

\usepackage{listings}
\usepackage{xcolor}

\definecolor{codegreen}{rgb}{0,0.6,0}
\definecolor{codegray}{rgb}{0.5,0.5,0.5}
\definecolor{codepurple}{rgb}{0.58,0,0.82}
\definecolor{backcolour}{rgb}{0.95,0.95,0.92}
\definecolor{bronze}{rgb}{0.8, 0.5, 0.2}
\definecolor{navyblue}{rgb}{0.0, 0.0, 0.5}
\definecolor{airforceblue}{rgb}{0.36, 0.54, 0.66}
\definecolor{blue}{rgb}{0.0, 0.0, 1.0}
\definecolor{blue(pigment)}{rgb}{0.2, 0.2, 0.6}
\lstdefinestyle{mystyle}{
    language=R,
    backgroundcolor=\color{backcolour},   
    commentstyle=\color{bronze},
    keywordstyle=\color{navyblue},
    numberstyle=\tiny\color{codegray},
    stringstyle=\color{codepurple},
    basicstyle=\ttfamily\footnotesize,
    breakatwhitespace=false,         
    breaklines=true,                 
    captionpos=b,                    
    keepspaces=true,                 
    numbersep=5pt,                  
    showspaces=false,                
    showstringspaces=false,
    showtabs=false,                  
    tabsize=2,    
    otherkeywords={!,!=,~},
    alsoother={.},
    morekeywords={stan_fit, optimal, plot_k_diagnostic, plot_posterior_diagnostic},
    classoffset=1, 
    keywordstyle=\color{airforceblue},
    otherkeywords={sfpca_data, Nsamples, Nchain, Ncores, PC_range, nknot_range,
    model_list},
    morekeywords={sfpca_data, Nsamples, Nchain, Ncores, PC_range, nknot_range,
    model_list},
    classoffset=2, 
    keywordstyle=\color{airforceblue},
    otherkeywords={<-, =},
    morekeywords={<-, =},
    classoffset=3, 
    keywordstyle=\color{blue},
    otherkeywords={0,1,2,3,4,5,6,7,8,9},
    morekeywords={0,1,2,3,4,5,6,7,8,9},
    classoffset=0
}

\begin{document}

\long\def\/*#1*/{} 

\def\spacingset#1{\renewcommand{\baselinestretch}%
{#1}\small\normalsize} \spacingset{1}


\if0\blind
{
  \title{\bf BayesTime: Bayesian Functional Principal Components for Sparse Longitudinal Data}
  \author[1]{Lingjing Jiang}
  \author[2]{Yuan Zhong}
  \author[3]{Chris Elrod}
  \author[1]{ Loki Natarajan}
  \author[4,5,6,7]{Rob Knight}
  \author[1]{Wesley K. Thompson}

  \affil[1]{\small Herbert Wertheim School of Public Health and Human Longevity Science, University of California San Diego, La Jolla, CA}
  \affil[2]{\small Department of Biostatistics, University of Michigan, Ann Arbor, MI}
  \affil[3]{\small Julia Computing, Boston, MA}
  \affil[4]{\small Department of Pediatrics, University of California San Diego, La Jolla, CA}
  \affil[5]{\small Center for Microbiome Innovation, University of California San Diego, La Jolla, CA}
  \affil[6]{\small Department of Computer Science and Engineering, University of California San Diego, La Jolla, CA}
  \affil[7]{\small Department of Bioengineering, University of California San Diego, La Jolla,CA}
  
  \/*\author{Lingjing Jiang, Yuan Zhong, Chris Elrod, \\
  Loki Natarajan, Rob Knight, Wes Thompson \thanks{
    The authors gratefully acknowledge \textit{please remember to list all relevant funding sources in the unblinded version}}\hspace{.2cm}\\
    Herbert Wertheim School of Public Health and Human Longevity Science, University of California, San Diego\\
    Department of Computer Science, University of California, San Diego}*/
    
  \maketitle
} \fi

\if1\blind
{
  \bigskip
  \bigskip
  \bigskip
  \begin{center}
    {\LARGE\bf Title}
\end{center}
  \medskip
} \fi

\bigskip
\begin{abstract} 
\noindent Modeling non-linear temporal trajectories is of fundamental interest in many application areas, such as in longitudinal microbiome analysis. Many existing methods focus on estimating mean trajectories, but it is also often of value to assess temporal patterns of individual subjects. Sparse principal components analysis (SFPCA) serves as a useful tool for assessing individual variation in non-linear trajectories; however its application to real data often requires careful model selection criteria and diagnostic tools. Here, we propose a Bayesian approach to SFPCA, which allows users to use the efficient leave-one-out cross-validation (LOO) with Pareto-smoothed importance sampling (PSIS) for model selection, and to utilize the estimated shape parameter from PSIS-LOO and also the posterior predictive checks for graphical model diagnostics. This Bayesian implementation thus enables careful application of SFPCA to a wide range of longitudinal data applications. 

\end{abstract}

\noindent%
\vfill

\newpage
\spacingset{1.5} 
\section{Introduction}
\label{sec:intro}
Longitudinal data, i.e., multiple observations collected on the same subject over time, are ubiquitous in biomedical research. In addition to using longitudinal data to estimate mean trajectories, it is often of great interest to characterize individual subject variation. Both the mean trajectory and individual subject deviations from the mean trajectory may be highly non-linear and hard to characterize using typical modeling approaches for longitudinal data such as linear mixed-effects models. Additionally, longitudinal data are often collected at irregular timing and frequency across subjects (they are ``sparse"), and methods for estimating trajectories need to be able to handle this common scenario.

For example, a question of fundamental interest in microbiome research is how the microbiome evolves in individual subjects as a response to subject-level perturbations, such as disease, diet and lifestyle \citep{kostic2015dynamics, halfvarson2017dynamics, smarr2017tracking, weingarden2015dynamic, david2014diet, turnbaugh2009effect, fierer2008influence}. Accurate continuous monitoring of a subject's microbiome may substantially improve prevention and treatment of some disorders. However, high-density temporal sampling is not currently feasible for microbiome studies; in practice, microbiome samples are collected infrequently and irregularly across time and subjects. Moreover, next generation sequencing techniques used to obtain estimates of microbial measurements are noisy, thus further hindering inference regarding the temporal evolution of a given subject's microbial status. Finally, the microbiome exhibits highly nonlinear dynamics over time, which introduces an additional complication to traditional longitudinal analysis methods. While several analytical methods have been developed to model microbiome temporal dynamics addressing these challenges \citep{ridenhour2017modeling, gibson2018robust, silverman2018dynamic, shenhav2019modeling, silverman2019bayesian}, by and large the focus has been on mean trajectories, substantially ignoring potentially important information about variation in trajectories across subjects. Since microbiome progression is highly idiosyncratic, it would be of great interest to capture relevant individual deviation from the mean trajectories, perhaps resulting in personalized predictions and clustering of subjects based on progression patterns.

Sparse functional principal components analysis (SFPCA) serves as a useful tool to estimate smooth mean trajectories while at the same time estimating smooth principal modes of variation of subject-level trajectories around the mean trajectory. SFPCA can be framed as an extension of linear random-effects models, where time effects are treated as random and non-linearity is achieved by choice of the functional basis \citep{james2000principal, kidzinski2018longitudinal}. The covariance structure of the trajectories is modeled as a low-rank matrix to produce efficient estimates of individual trajectories. Various fitting approaches, such as the EM algorithm, kernel smoothing and Newton-Raphson algorithm, have been proposed to estimate parameters of the SFPCA model \citep{james2000principal, yao2005functional, peng2009geometric}. These approaches then use model selection techniques, such as cross-validation, Akaike information criterion (AIC) and leave-one-curve-out cross-validation, to select the dimension of basis and the number of principal components. However, due to their reliance on assumptions such as normally-distributed component scores and residuals, these models need to be carefully examined when applied to real data \citep{kidzinski2018longitudinal}. 

We implemented the SFPCA model in a Bayesian framework to provide a flexible modeling approach that incorporates effective model selection and graphical diagnostic methods. Our \textsf{BayesTime} R package implementing the Bayesian SFPCA model allows users to use leave-one-out cross-validation (LOO) with Pareto-smoothed importance sampling (PSIS) for model selection \citep{vehtari2017practical}, and to utilise the estimated shape parameter from PSIS-LOO and graphical posterior predictive checks for model diagnostics \citep{gelman1996posterior, gabry2019visualization}. This Bayesian implementation thus offers a flexible and comprehensive solution to real-date SFPCA applications,  such as longitudinal microbiome data.

The Bayesian framework of SFPCA with PSIS-LOO is described in Section 2, and is implemented in the \textsf{BayesTime} package in R (Section 3). Section 4 presents Monte Carlo simulations evaluating the Baysian SFPCA model performance and further illustrates its use on a real longitudinal microbiome dataset, showing how individual microbiome trends can be visualized and explored with \textsf{BayesTime}. Future work is discussed in Section 5.

\section{Methods}
\label{sec:meth}
\subsection{Sparse Functional Principal Components Analysis}
The classical assumption of functional data analysis is that each trajectory is sampled over a dense grid of time points common to all individuals \citep{ramsay2007applied}. However, in practice, trajectories are often measured at an irregular and sparse set of time points that can differ widely across individuals. To address this scenario, \citet{james2000principal} proposed {\it sparse functional principal components analysis} (SFPCA) using a reduced rank mixed-effects framework. Let $Y_i (t)$ be the measurement at time $t$ for the $i$th individual, $\mu(t)$ the overall mean function, $f_j$ the $j$th principal component function and $\boldsymbol{f}= [(f_1,f_2,…,f_k)]^T$, where $k$ is the number of principal components. Then the \citet{james2000principal} SFPCA model is given by 
\begin{equation}
    Y_i(t) = \mu(t) + \sum_{j=1}^{k}f_j(t) \alpha_{ij} + \epsilon_i(t), \quad i = 1, ..., N
\end{equation}
subject to the orthogonality constraint $\int f_j  f_l = \delta_{jl}$, the Kronecker $\delta$. The vector 
$\boldsymbol \alpha_i = (\alpha_{i1},\ldots,\alpha_{ik})^T$ is the component weights for the $i$th individual and $\epsilon_i (t)$ is a normally-distributed residual, independent across subjects and across times within subject. The functions $\mu$ and $\boldsymbol{f}$ are approximated using cubic splines to allow for a smooth but flexible fit. Let $\boldsymbol b(t)$ be a cubic spline basis with dimension $q > k$.
The spline basis is orthonormalized so that $\int b_j b_l = \delta_{jl}$.
Let $\Theta$ and $\boldsymbol\theta_\mu$ be, respectively, a $q \times k$ matrix and a $q$-dimensional vector of real-valued coefficients. For each individual $i$, denote their measurement times by $\boldsymbol t = (t_{i1},t_{i2},…,t_{in_i})^T$, and let  $\boldsymbol Y_i= (Y_i (t_{i1}),…,Y_i (t_{in_i }))^T$ be the corresponding real-valued observations. Then $B_i= (\boldsymbol b(t_{i1}),…,\boldsymbol b(t_{in_i}))^T$ is the $n_i \times q$ spline basis matrix for the $i$th individual. The reduced rank model can then be written as 
\begin{align}
    Y_i = B_i \boldsymbol\theta_\mu + B_i \Theta \boldsymbol\alpha_i + \epsilon_i, \quad i = 1, ..., N, \label{eq:fpca} \\
    \Theta^T \Theta = I, \quad \boldsymbol\alpha_i \sim N(0, D), \quad \epsilon_i \sim N(0, \sigma_\epsilon^2 I_{n_i}), \nonumber 
\end{align} 
where the covariance matrix $D$ is restricted to be diagonal and $I_{n_i}$ is the $n_i \times n_i$ identity matrix.

\subsection{Bayesian SFPCA}
We developed the SFPCA model in a Bayesian framework to allow for flexible prior specification and implementation of model selection and assessment methods. We implemented this Bayesian model using Hamilton Markov Chain Monte Carlo (MCMC) sampling algorithm in \textsf{Stan} \citep{carpenter2017stan}. The real-valued observations $Y_i(t)$ are first standardized to have mean zero and standard deviation one. The prior distributions for parameters in Eq. (\ref{eq:fpca}) were chosen as follows:
\begin{align*}
    \boldsymbol \theta_\mu  &\sim N_q (0,I_q ) \\
    \boldsymbol \alpha_i &\sim N_k (0,I_k ) \\
    \boldsymbol \Theta_j &\sim N_q (0,I_q ), j=1,\ldots,k \\
    \epsilon_i  &\sim N_{v_i} (0,\sigma_\epsilon^2 I_{v_i}) \\
    \sigma_\epsilon &\sim Cauchy(0,1),
\end{align*}
where $\boldsymbol \Theta_j$ is the $j$th column of the loading matrix $\Theta$, and $v_i$ is the total number of visits for the $i$th subject. The Bayesian implementation also enables use of 
leave-one-out cross-validation with Pareto-smoothed important sampling (PSIS-LOO) \citep{vehtari2017practical} to perform model selection on the number of principal components $k$ 
and the number of basis functions $q$.  Moreover, model fit can be assessed via diagnostics plots from PSIS-LOO as well as the graphical posterior predictive checks obtained from simulating posterior predictive data \citep{gelman1996posterior, gabry2019visualization}.

One difficulty in implementing the Bayesian SFPCA model is that the principal component loadings $\Theta$ are not uniquely specified. 
For a given $k \times k$ rotation matrix $P$, if $\Theta^* = \Theta P$ and $\Theta$ obeys the constraints in Eq.(\ref{eq:fpca}), then $\Theta^{*T} \Theta^*= P^T \Theta^T \Theta P = I$, and hence $\Theta$ is unidentifiable without additional restrictions. Instead of directly enforcing orthonormality when sampling from the conditional posteriors in the Bayesian model fitting, we sampled the parameters with no constraint on $\Theta$ and then performed a {\it post hoc} rotation for each 
iteration of the MCMC algorithm to meet the orthonormality constraint.
Since the symmetric matrix $\Theta^T \Theta$ is identifiable and non-negative definite, we applied an eigenvalue decomposition $\Theta^T \Theta = V S V^T$, where $V$ is the $q \times q$ matrix of orthonormal eigenvectors, and $S$ is the diagonal matrix of eigenvalues, with the $q$ positive eigenvalues ordered from largest to smallest. Let $\Theta^* = V_k$ denote the $q \times k$ matrix consisting of the first $k$ eigenvectors of $V$, which satisfies $\Theta^{*T} \Theta^{*} = I$. 
Finally, we rotated the FPC scores $\boldsymbol\alpha_i$ to obtain $\boldsymbol \alpha_i^* = \Theta^{*T} \Theta \boldsymbol\alpha_i$, so that $\Theta^* \boldsymbol\alpha_i^{*} = \Theta \boldsymbol\alpha_i$.

\subsection{Model Selection with PSIS-LOO}
Leave-one-out cross-validation(LOO) with Pareto smoothed importance sampling (PSIS) is a stable model selection procedure which  has been shown to be more robust in the presence of influential observations than other widely used criteria such as AIC (Akaike information criterion), DIC (deviance information criterion) and WAIC (widely applicable information criterion) \citep{vehtari2017practical}. 
In Bayesian leave-one-out cross-validation, the estimate of the out-of-sample predictive fit (expected log pointwise predictive density) is defined as 
\begin{equation}
    elppd_{loo} = \sum_{i=1}^n \log p(y_i | y_{-i}),
\end{equation}
where $p(y_i|y_{-i}) = \int p(y_i | \theta) p(\theta|y_{-i}) d\theta$ is the leave-one-out predictive density given the data without the $i$th data point.
Typically, LOO-CV requires re-fitting the model $n$ times; however, a computational shortcut exists to enable only one model evaluation. As noted by
\citet{gelfand1992model}, if the $n$ points are conditionally independent in the data model, we can then evaluate $p(y_i | y_{-i})$ with
draw $\theta^s$ from the full posterior $p(\theta|y)$ using importance ratios, defined as 
\begin{equation}
    r_i^s =  \frac{1}{p(y_i |\theta^s)}  \propto  \frac{p(\theta^s |y_{-i})} {p(\theta^s |y)}.
\end{equation}
However, the posterior $p(\theta|y)$ is likely to have a smaller variance and thinner tails than the leave-one-out distribution
$p(\theta|y_{-i})$, and thus a direct use of the formula above induces instability, because the importance ratios can have high or infinite variance. 

\citet{vehtari2015pareto} improve the LOO estimate using Pareto smoothed importance sampling (PSIS), which applies a smoothing procedure to the importance weights. As the distribution of the importance weights used in LOO may have a long right tail, the empirical Bayes estimate of \citet{zhang2009new} can be used to fit a generalized Pareto distribution to the tail (e.g. $20\%$ largest importance rations), and this is done separately for each held-out data point $i$. So for each $i$, the result is a vector of weights $\widetilde{w_i}^s = F^{-1} \left( \frac{z-\frac{1}{2}}{M} \right), z = 1, …, M$, where $M$ is the number of simulation draws used to fit the Pareto distribution (in this case, $M=0.2S$), and $F^{-1}$ is the inverse-CDF of the generalized Pareto distribution. Then each vector of weights is truncated at $S^{3/4} \bar{w_i}$, denoted as $w_i^s$. These results can then be combined to compute the PSIS estimate of the LOO expected log pointwise predictive density: 

\begin{equation}
    \widehat{elppd _{psis-loo}} = \sum_{i=1}^n \log \frac{\sum_{s=1}^S w_i^s  p(y_i |\theta^s)} {\sum_{s=1}^S w_i^s}.
\end{equation}

\subsection{Model Diagnostics}
PSIS-LOO is not only efficient, it can also provide useful diagnostics for model checking. The estimated shape parameter $\hat k$ of the fitted Pareto distribution can be used to assess the reliability of the estimate; this diagnostic approach can be used routinely with PSIS-LOO for any model with a factorizable likelihood. If $ k < \frac{1}{2}$, the variance of the raw importance ratios is finite, the central limit theorem holds, and the estimate converges quickly. If $k \in [\frac{1}{2},1] $, the variance of the raw importance ratios is infinite but the mean exists, the generalized central limit theorem for stable distributions holds, and the convergence of the estimate is slower. If $k>1$, the variance and the mean of the raw ratios distribution do not exist. \citet{vehtari2017practical} suggested that if the estimated tail shape parameter $\hat{k}$ exceeds 0.5, the user should be warned, although in practice they have observed good performance of values of $\hat{k}$  up to 0.7. Hence, this threshold of 0.7 could be used in practice for model diagnostics. If the $i$th LOO predictive distribution has a large $\hat k$ value when holding out data point $i$ to evaluate predictive density, it suggests that data point $i$ is a highly influential observation that deserves further examination.

Moreover, since we are implementing a Bayesian SFPCA  model, we can also  compare the observed data to simulated data from the posterior predictive distribution \citep{gabry2019visualization}. The idea behind posterior predictive checks is simple: if a model is a good fit, then it should be able to generate data that resemble the observed data. The data used for posterior predictive checks are simulated from the posterior predictive distribution $p(\tilde y|y) = \int p(\tilde y|\theta) p(\theta|y) d\theta$, where $y$ is the current observed data, $\tilde y$ is the new data to be predicted, and $\theta$ are model parameters. By comparing the observed and replicated data in the posterior predictive checks, we may find a need to extend or modify the model.

\section{Implementation}
\label{sec:implement}
The Bayesian SFPCA method has been implemented in R in the \textsf{BayesTime} package at \emph{github.com/biocore/bayestime}. The user can choose a range of the number of principal components and the dimension of cubic spline basis (i.e. the number of internal knots + 4) by the \textsf{PC{\_}range} and \textsf{nknot{\_}range} argument in \textsf{stan{\_}fit} function, with which the knots are placed by the quantile of the time range in the default setting. The following model comparisons are performed with the \textsf{optimal{\_}model} function, which compare models based on their $\widehat{elppd_ {psis-loo}}$ and standard errors. Moreover, model diagnostics on the chosen model can be visualized using \textsf{plot{\_}k{\_}diagnostic} and \textsf{plot{\_}posterior{\_}diagnostic} fuctions. 

\begin{lstlisting}
library(BayesTime)
# use PSIS-LOO for model selection
sfpca_stan_results <- stan_fit(sfpca_data = dat, Nsamples = 1000, Nchain = 3, Ncores=3, PC_range = c(1,2,3), nknot_range = c(1,2))   
optimal_model_idx <- optimal(model_list = sfpca_stan_results)
optimal_model <- sfpca_stan_results[[optimal_model_idx]]
# model diagnostics with Pareto shape parameter k 
plot_k_diagnostic(dat, optimal_model)
# model diagnostics with posterior predictive checking 
plot_posterior_diagnostic(dat, optimal_model) 

\end{lstlisting}

\section{Examples}
\label{sec:examples}
Using the \textsf{BayesTime} package, we evaluated the performance of the Bayesian SFPCA model in Monte Carlo simulation studies and applied it to a longitudinal microbiome dataset to demonstrate its utility in a practical example. Data and code for simulations and real data application are available at \emph{https://github.com/knightlab-analyses/BayesTime-analyses}.

\subsection{Simulation Studies}
Due to potential sequencing errors and sample collection procedures, missing data and dropouts are the norm rather than the exception in longitudinal microbiome studies. Moreover, despite large-scale cross-sectional microbiome studies such as the Human Microbiome Project \citep{turnbaugh2007human, methe2012framework} and American Gut Project \citep{mcdonald2018american}, studies characterizing human-associated microbial communities over time often have relatively small sample sizes \citep{dethlefsen2011incomplete, flores2014temporal, caporaso2011moving}. Hence, it is important to assess the performance of Bayesian SFPCA in simulations with various sample sizes and with different levels of sparsity. In our simulations, we varied the total number of subjects at 100, 50, 25, 10, and the proportion of missing values at 0\%, 20\%, 50\% and 80\% (i.e., the percentage of randomly deleted observations to create increasingly sparse functional datasets) over observations at 10 time points. To better mimic the reality, we simulated longitudinal trajectories based on an SFPCA model using parameters initially estimated from the real microbiome data in the following way: 

\begin{enumerate}
    \item applying SFPCA to a real longitudinal microbiome dataset~\citep{dominguez2016partial}; 
    \item selecting the optimal number of PCs $k$ and dimension of basis $q$ using PSIS-LOO;
    \item extracting the estimated values for population mean curve $(\theta_\mu)$, FPC loadings $(\Theta)$,  diagonal covariance matrix of FPC scores $(D)$, and error variance $(\sigma^2)$.
\end{enumerate}

Then we simulate the data by varying the number of subjects, the number of time points and proportion of missing data as follows:
\begin{enumerate}
    \item 	choosing the total number of subjects (N) and of time points $(N_T)$ in order to place possible time points between $[0,1]$;
    \item  specifying the average number $(\mu_T)$ of time points across all subjects in order to vary the proportion of missing data (approximated as $1-\mu_T/N_T$) by simulating the observed number of time points for each individual with $n_i \sim Poisson (\mu_T)$ and then randomly placing the observed time points in the possible time locations (chosen in the previous step);
    \item  generating the cubic spline basis matrix $b(t)$ for each subject (orthonormality obtained through Gram-Schmidt orthonormalization); 
    \item  simulating for each subject FPC scores $\alpha_i \sim N(0,D)$ and noise $\epsilon_i  \sim N(0,\sigma^2 I)$;
    \item 	obtaining the temporal trajectory for each individual with $Y_i (t)= B_i \theta_\mu+ B_i \Theta \alpha_i+ \epsilon_i$, where $B_i=(b(t_{i1}),…, b(t_{in_i}))^T$
    \item repeating steps 1 -- 5 100 times for each simulation scenario with different number of subjects and proportion of missing data, thus generating 1600 simulated datasets in total.
\end{enumerate}

\begin{figure}
\begin{center}
\includegraphics[width=5.5in]{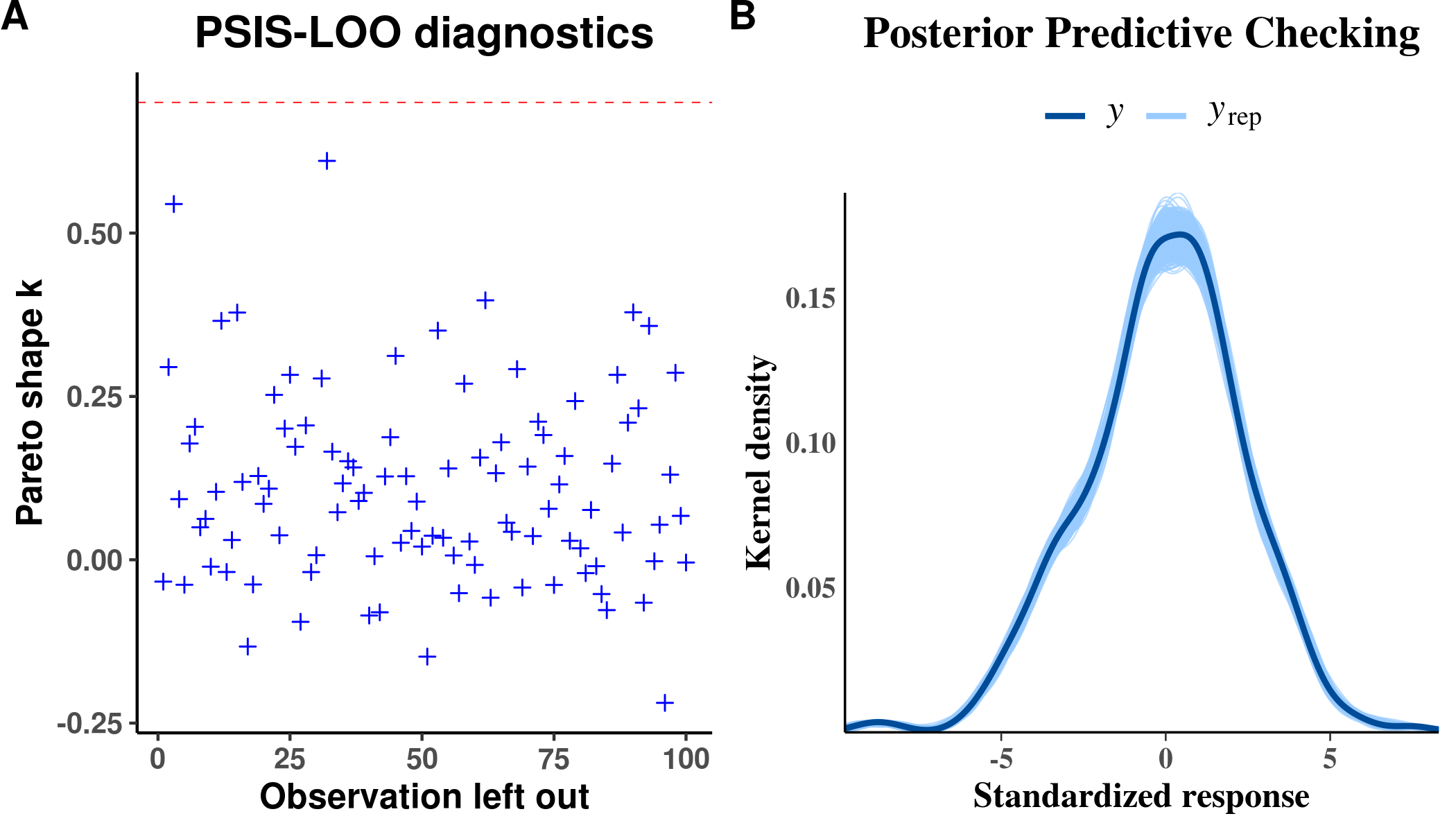}
\end{center}
\caption{Graphical model checking with PSIS-LOO diagnostic plot and posterior predictive checks for Bayesian SFPCA simulated scenario of 100 subjects with 80\% missing data. (A) Scatterplot of estimated Pareto shape parameter $\hat k$ in PSIS-LOO diagnostic plot: all $\hat k$'s are lower than the warning threshold 0.7. (B) Graphical posterior predictive checks: kernel density estimate of the observed dataset $y$ (dark curve), with kernel estimates for 100 simulated datasets $y_{rep}$ drawn from the posterior predictive distribution (thin, lighter lines). \label{sim_fig0}}.
\end{figure}

Before describing the simulation results, we want to use the scenario of 100 subjects with 80\% missing data to demonstrate how to perform model selection with PSIS-LOO and how to use its estimated shape parameter $\hat k$ to assess the reliability of the model. Models are compared based on their values of $\widehat{elppd _{psis-loo}}$: the larger the value, the better the model is. Among nine models tested (with the number of PCs and the number of internal knots ranging from 1 to 3), the model with two PC's and one internal knot had the highest $\widehat{elppd _{psis-loo}}$. The second best model (with three PC's and one internal knot) is lower in $\widehat{elppd _{psis-loo}}$ by 1.86, and the standard error of the difference between the two models is 2.21, indicating that the second model provides a similarly good fit. But since the first model is more parsimonious and all of its estimated shape parameters $\hat k$ are smaller than 0.7 (Figure~\ref{sim_fig0}A), we chose this as our best model. 

We also generated graphical displays comparing observed data to simulated data from the posterior predictive distribution. In Figure~\ref{sim_fig0}B, the dark line is the distribution of the observed outcomes $y$ and each of the lighter lines is the kernel density estimate of one of the replicates of $y$ from the posterior predictive distribution. This figure shows that there is very little discrepancy between real and simulated data from the model, confirming the model validity for this application. 

\begin{figure}
\begin{center}
\includegraphics[width=5.5in]{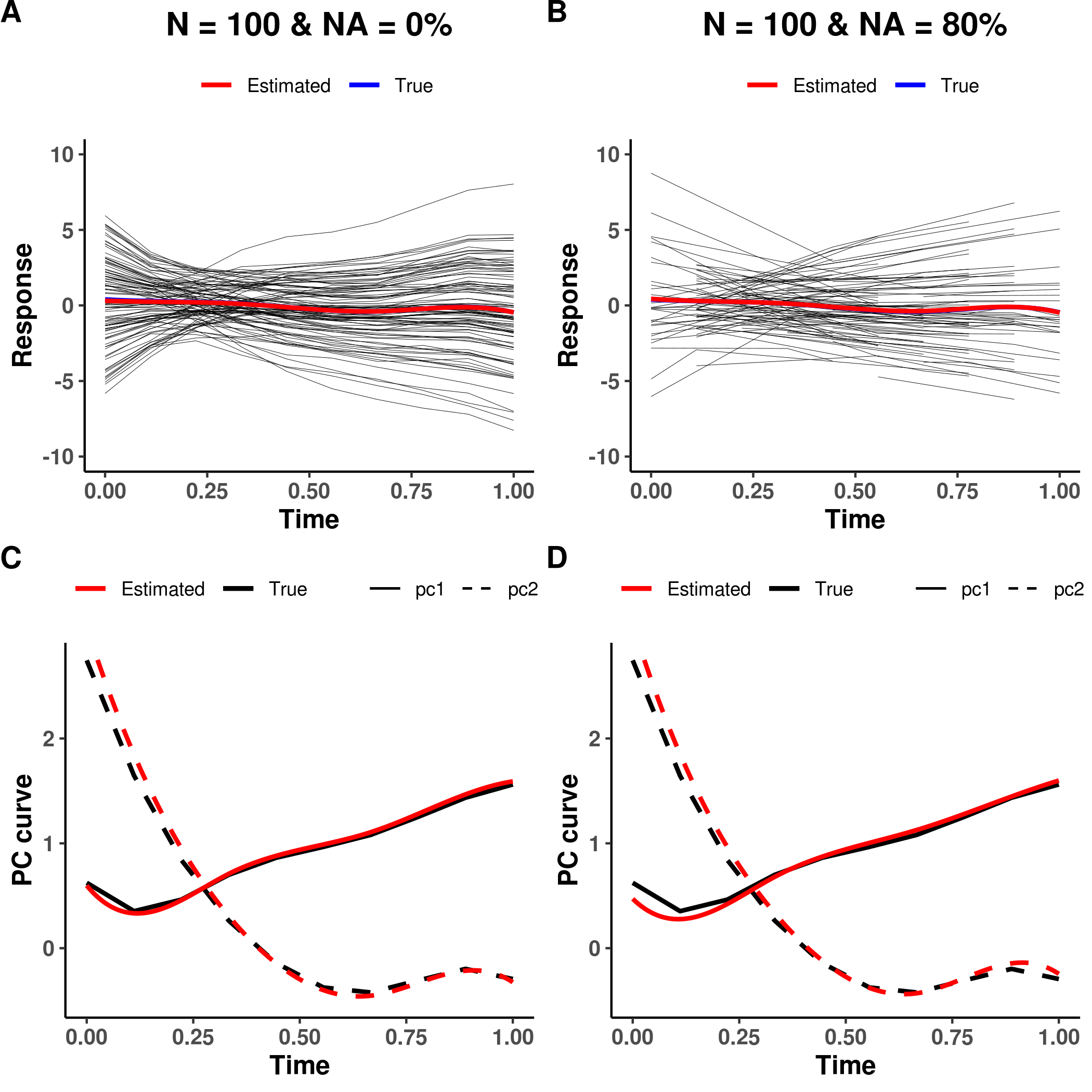}
\end{center}
\caption{Results of Bayesian SFPCA on simulated data with 100 total samples of 0\% vs. 80\% missing data. (A) Estimated (red) vs. true (blue) overall mean curve on simulated trajectories (black) with 0\% missing data. (B) Estimated (red) vs. true (blue) overall mean curve on simulated trajectories (black) with 80\% missing data. (C) Estimated (red) vs. true (black) PC curves on simulated data with 0\% missing data. (D) Estimated (red) vs. true (black) PC curves on simulated data with 80\% missing data. \label{sim_fig1}}.
\end{figure}

\begin{figure}
\begin{center}
\includegraphics[width=5.5in]{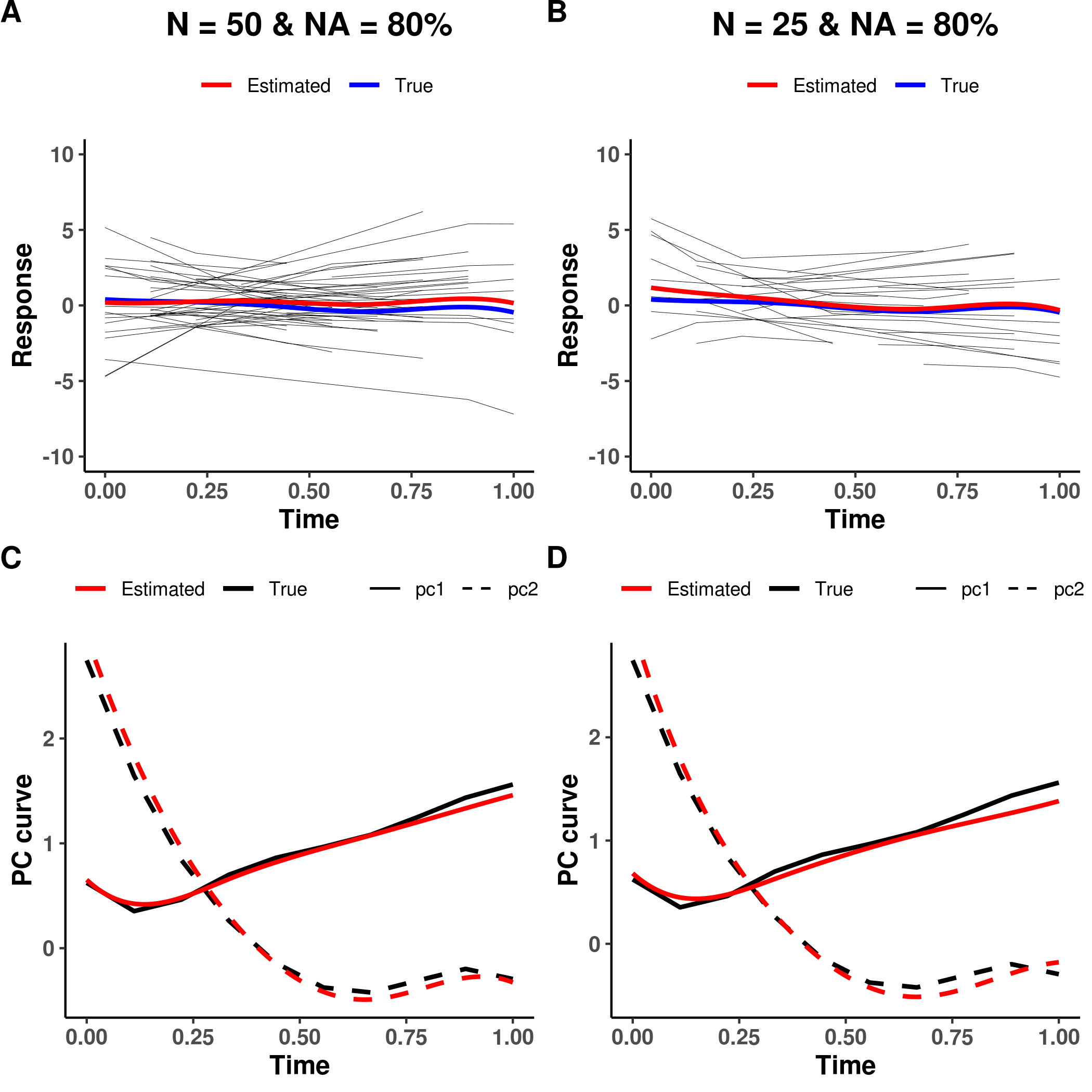}
\end{center}
\caption{Results of Bayesian SFPCA on simulated data with 50 vs. 25 total samples of 80\% missing data. (A) Estimated (red) vs. true (blue) overall mean curve on simulated trajectories (black) with 50 samples. (B) Estimated (red) vs. true (blue) overall mean curve on simulated trajectories (black) with 25 samples. (C) Estimated (red) vs. true (black) PC curves on simulated data with 50 samples. (D) Estimated (red) vs. true (black) PC curves on simulated data with 25 samples.  \label{sim_fig2}}.
\end{figure}

\begin{figure}
\begin{center}
\includegraphics[width=5.5in]{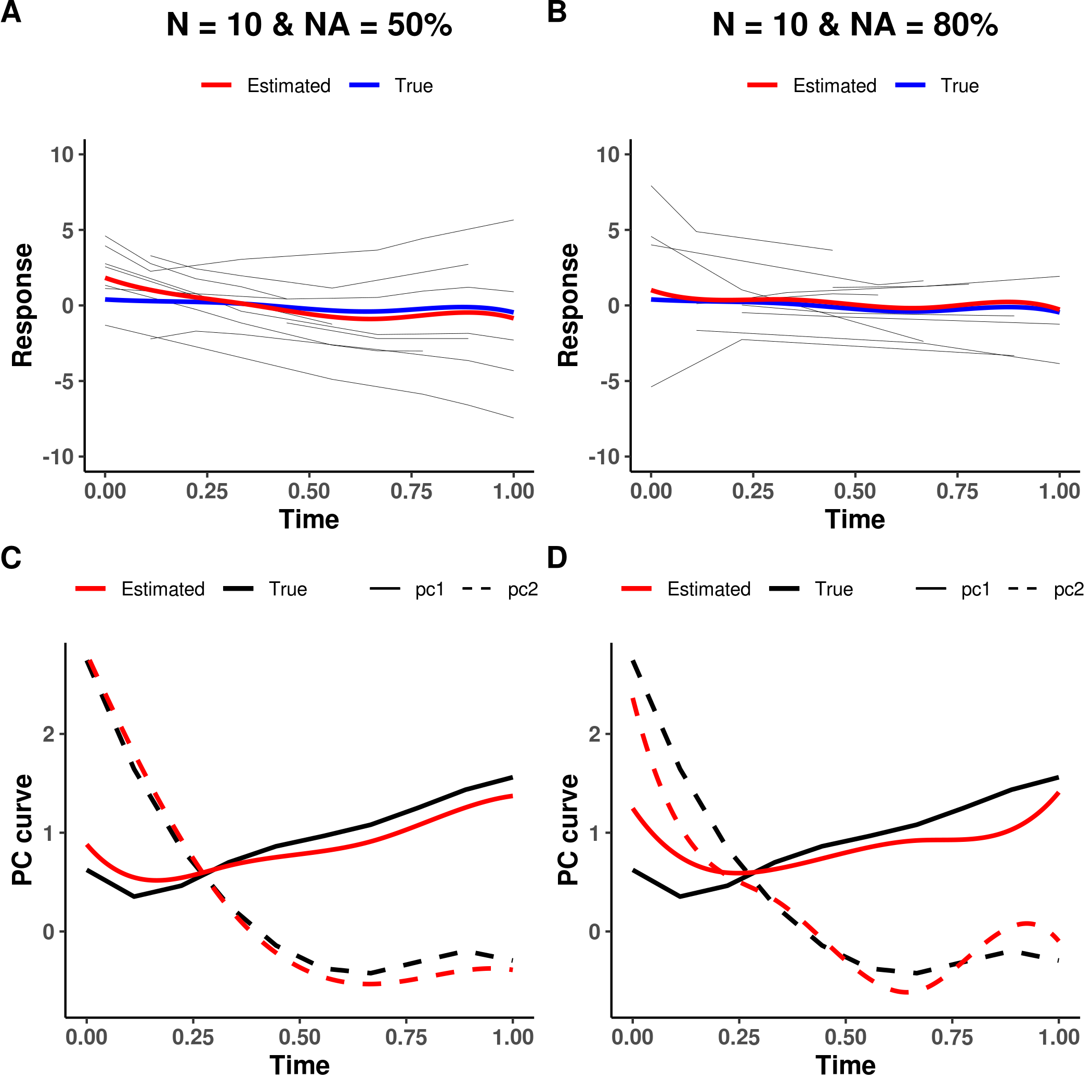}
\end{center}
\caption{Results of Bayesian SFPCA on simulated data with 10 total samples of 50\% vs. 80\% missing data. (A) Estimated (red) vs. true (blue) overall mean curve on simulated trajectories (black) with 50\% missing data. (B) Estimated (red) vs. true (blue) overall mean curve on simulated trajectories (black) with 80\% missing data. (C) Estimated (red) vs. true (black) PC curves on simulated data with 50\% missing data. (D) Estimated (red) vs. true (black) PC curves on simulated data with 80\% missing data. \label{sim_fig3}}.
\end{figure}

To evaluate the performance of Bayesian SFPCA, we investigated how well it recovered the mean trajectory and two PC functions. With 100 subjects, even as the proportion of missing data increased from 0\% to 80\%, the estimated overall mean curves and PC curves accurately recovered the ground truth in both scenarios (Figure~\ref{sim_fig1}). For the scenarios with 50 or 25 subjects with 80\% missing data, the estimated mean curves were still close to the ground truth, except for a slight deviation at the two ends due to the large proportion of missing data there (Figure~\ref{sim_fig2}A, B). The PC curves were estimated well for both cases on two PCs, despite slight underestimation toward the end on both PCs (Figure~\ref{sim_fig2}C, D). As for the challenging scenarios of 10 samples with 50\% or 80\% missing data, the estimated mean curves in both scenarios and the PC curves for the scenario with 50\% missingness were still robust (Figure~\ref{sim_fig3}A, B, C). However, for the case with 80\% missing data, the estimated PC 1 curve did not capture the decreasing trend as accurately as before and displayed an artificial curvature towards the end; the estimated PC 2 curve also exhibited some deviations from the ground truth (Figure~\ref{sim_fig3}D). A closer look at the simulated trajectories (Figure~\ref{sim_fig3}B) indicated that few trajectories exhibited the decreasing trend at the beginning in both PCs due to the loss of data, hence the deviated estimation was caused by the limitation of the underlying data. Note that the visual comparisons above were demonstrated using one representative case from each simulation scenario. Results over all 100 simulated datasets for each scenario were summarized in table~\ref{table_mean} and ~\ref{table_fpc}, showing that in the estimations of both mean ($\theta_\mu$) and FPC spline coefficients ($\Theta$), mean squared errors increase as sample size decreases at each given missing proportion, although the variabilities are still within the 95\% credible intervals. Moreover, errors for the mean and FPC estimations remain similar despite increasing missing proportion at fixed sample size. In summary, the performance of Bayesian SFPCA is robust to limited sample size and a high proportion of missing data.

\begin{table}
\caption{Average mean squared errors with 95\% CIs for estimating mean spline coefficients ($\theta_\mu$) in simulations with different sample size ($N$) and proportion of missing values ($NA$).  \label{table_mean}}
\begin{tabular}{ccccc}
 & N = 100 & N = 50 & N = 25 & N = 20 \\\hline
NA = 0\% & \makecell{0.008 (0, 0.029)}   & \makecell{0.015 (0, 0.066)} & \makecell{0.031 (0, 0.173)} & \makecell{0.044 (0.002, 0.188)}\\
NA = 20\% &  \makecell{0.008 (0, 0.034)} &  \makecell{0.01 (0, 0.052)} &  \makecell{0.021 (0.001, 0.088)} &  \makecell{0.04 (0.001, 0.138)}\\
NA = 50\% &  \makecell{0.008 (0, 0.034)} & \makecell{0.011 (0, 0.053)} & \makecell{0.021 (0.001, 0.082)} & \makecell{0.041 (0.001, 0.154)} \\
NA = 80\% &  \makecell{0.007 (0, 0.025)} & \makecell{0.013 (0.001, 0.08)} & \makecell{0.026 (0.001, 0.105)} & \makecell{0.069 (0.009, 0.223)} \\
\end{tabular}
\end{table}

\begin{table}
\caption{Average mean squared errors with 95\% CIs for estimating FPC spline coefficients ($\Theta$) in simulations with different sample size ($N$) and proportion of missing values ($NA$).  \label{table_fpc}}
\begin{tabular}{ccccc}
 & N = 100 & N = 50 & N = 25 & N = 20 \\\hline
NA = 0\% & \makecell{0.002 (0.001, 0.01)} & \makecell{0.011 (0.001, 0.116)} & \makecell{0.02 (0.001, 0.143)} & \makecell{0.049 (0.003, 0.21)} \\
NA = 20\% & \makecell{0.002 (0.001, 0.005)} & \makecell{0.005 (0.001, 0.03)} & \makecell{0.018 (0.001, 0.157)} & \makecell{0.05 (0.003, 0.208)} \\
NA = 50\% & \makecell{0.002 (0.001, 0.005)} & \makecell{0.005 (0.001, 0.026)} & \makecell{0.02 (0.001, 0.162)} & \makecell{0.054 (0.003, 0.203)} \\
NA = 80\% & \makecell{0.004 (0.001, 0.016)} & \makecell{0.01 (0.001, 0.079)} & \makecell{0.027 (0.002, 0.208)} & \makecell{0.081 (0.017, 0.246)} \\
\end{tabular}
\end{table}

\subsection{Impact of Skin Care Products on Microbiome Dynamics}
In this example, researchers want to know how the skin microbiome would be altered when the hygiene routine is modified, and whether this alteration is similar across different body sites \citep{bouslimani2019impact}. Twelve healthy subjects participated in this 9-week study, and samples were collected from each individual on four skin body sites (face, armpits, front forearms and toes). For the baseline (week 0), subjects performed  their normal routine of using their personal skin care products. During the first three weeks (w1-w3), all volunteers used only the same head-to-toe shampoo and no other product was applied. In the following 3 weeks (w4-w6), four selected commercial products were applied daily by all volunteers on the specific body site: sunscreen for the face, deodorant antiperspirant for the armpits, moisturizer for the front forearm, and soothing foot powder for the toes,
and continued use of the same shampoo. For the last three weeks (w7-w9), all volunteers went back to their normal routine using their personal products. Due to its specific study design, the perturbations of the skin microbiome are expected to occur around the intervention time points. The outcome of interest in this example is the longitudinal pattern of Shannon microbial diversity of the microbiome, defined as $Shannon = - \sum_{i=1}^{S} p_i ln(p_i) $, where $S$ is the total number of species, and $p_i$ is the relative proportion of species $i$ relative to the entire population.

\begin{figure}
\begin{center}
\includegraphics[width=5.5in]{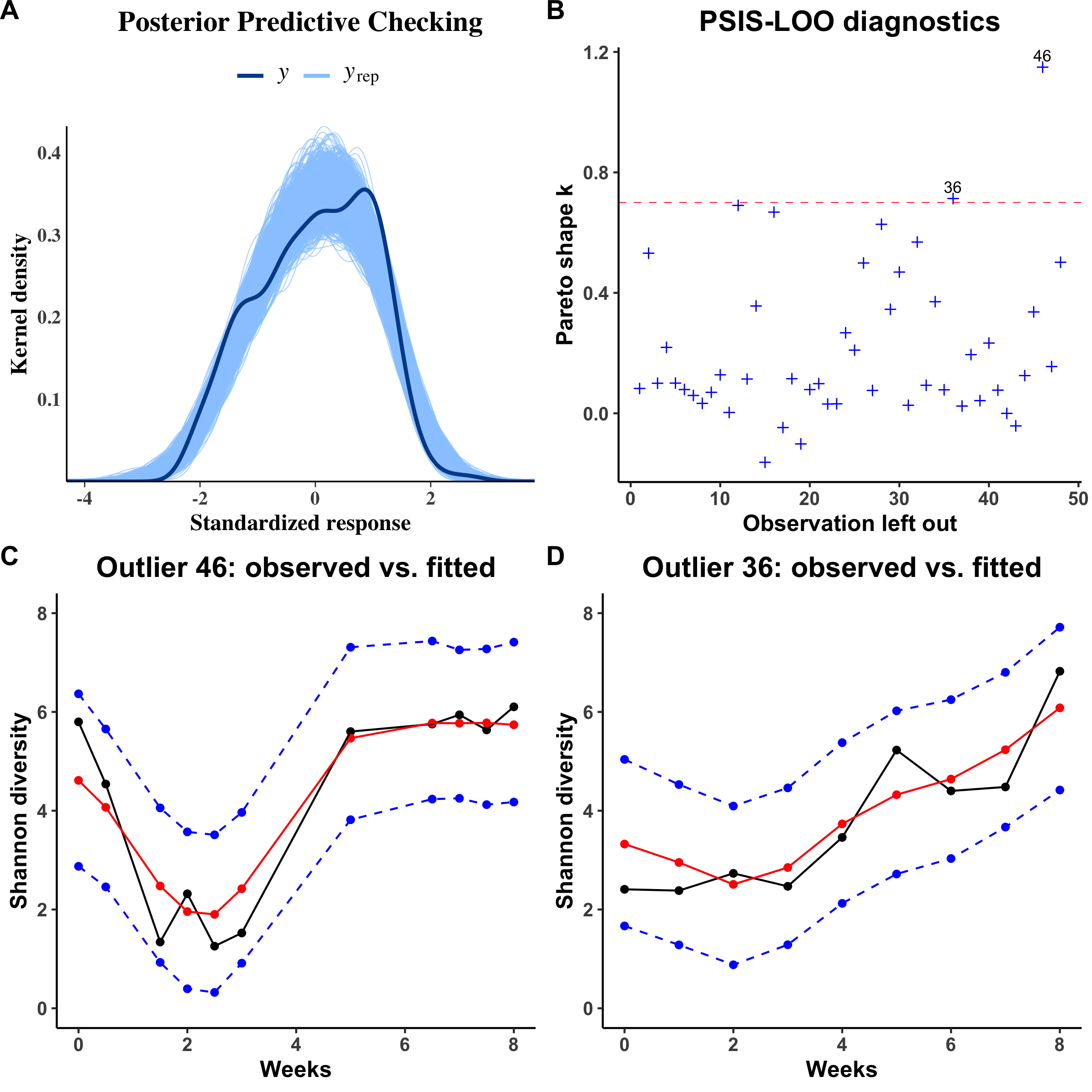}
\end{center}
\caption{Graphical model diagnostics with posterior predictive checks and PSIS-LOO diagnostic plot for Bayesian SFPCA application to skin microbiome dataset. (A) Graphical posterior predictive checks: kernel density estimate of the observed dataset $y$ (dark curve), with kernel estimates for 100 simulated dataset $y_{rep}$ drawn from the posterior predictive distribution (thin, lighter lines). (B) Scatterplot of estimated Pareto shape parameter $\hat k$ in PSIS-LOO diagnostic plot: all but two $\hat k$'s are lower than the warning threshold 0.7. (C) Observed (black), predicted (red) trajectories with 95\% credible intervals (dashed blue lines) for potential outlier 46. (D) Observed (black), predicted (red) trajectories with 95\% credible intervals (dashed blue lines) for potential outlier 36. \label{fig_skin_1}}
\end{figure}

 The best SFPCA model was selected by PSIS-LOO to have four PCs and three internal knots for the cubic spline basis. The estimated difference of expected leave-one-out prediction errors between the models with three and four PCs was smaller than the standard error, hence they could both be considered as adequate models. We chose the model with the highest value of $\widehat{elppd _{psis-loo}}$, which has four principal components and three internal knots. The model diagnostics using graphical posterior predictive checks showed that the simulated data from the posterior predictive distribution was able to cover the distribution of observed outcomes well (Figure~\ref{fig_skin_1}A). Moreover, the estimated shape parameters from PSIS-LOO were all under the threshold of 0.7, except for one subject with a marginal value at 0.71 and another with an extreme value at 1.15 (Figure~\ref{fig_skin_1}B). To examine these two potential outliers, we compared their observed trajectories with the predicted curves. Figure~\ref{fig_skin_1}C, D showed that the observed trajectories (black) were closely followed by the predicted curves (red) and fell within the 95\% credible intervals (blue). All these suggested that our selected SFPCA model was able to fit this dataset well. 

\begin{figure}
\begin{center}
\includegraphics[width=5in]{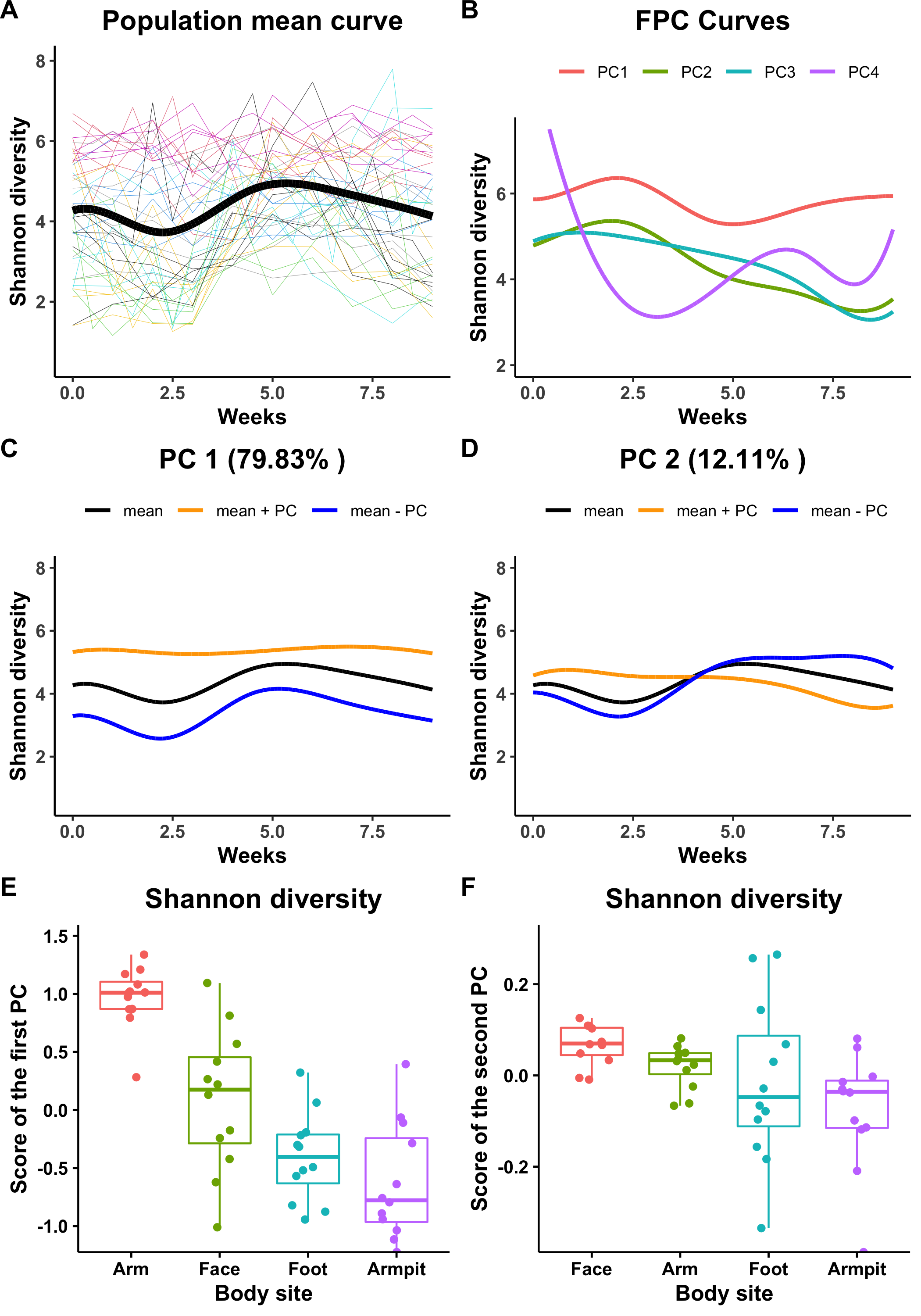}
\end{center}
\caption{Results of Bayesian SFPCA on skincare impact microbiome dataset. (A) Estimated population mean curve (black) on observed individual trajectories (colored). (B) Trends of variability captured by 4 principal component curves. (C) Effects of adding (orange) and subtracting (blue) PC 1 with 1 SD of PC score to the population mean curve (black). (D) Effects of adding (orange) and subtracting (blue) PC 2 with 1 SD PC score to the population mean curve (black). (E) Progression trends based on PC 1 scores in different body sites. (F) Progression trends based on PC 2 scores in different body sites. \label{fig_skin_2}}.
\end{figure}

As seen in Figure~\ref{fig_skin_2}A, the population mean curve reveals an overall trend of an initial decrease in microbial diversity during the first 3 weeks due to the cessation of using personal skin care products, an increase in the middle 3 weeks because of the introduction of four additional products, and a decrease toward the end due to the resumption of normal routines. Figure~\ref{fig_skin_2}B shows the first four estimated PCs, with the first two PCs explaining over 90\% of the variance. The first principal component captures variation in  changes in microbial diversity around week 2.5 and week 5. The second component captures additional variation in changes around week 8. In Figure~\ref{fig_skin_2}C-D, by adding a PC with $\pm 1$ standard deviation of PC scores to the population mean curve, we illustrate how the first and second PCs impact the trajectories. The first PC represents an overall vertical shift of the mean microbial diversity, and explains about 80\% of the variance. An individual with a high score on this component has on average higher microbial diversity than one with a lower score, and {\it vice versa}. The second PC curve explains $12\%$ of the variance, and captures variation during the middle three weeks. Since a trajectory of each individual is represented as a weighted sum of these principal modes of variation, we can use each individual's PC scores to gain insight about microbial perturbations in different body sites (Figure~\ref{fig_skin_2}E-F). The scores of the first PC unveil the order of microbial diversity from highest to lowest in the four body sites, where arm has the highest diversity over time, while armpit the lowest. The signs of mean scores (positive or negative) indicate that arm and face share one similar temporal pattern, corresponding to the orange curve in Figure~\ref{fig_skin_2}C, while foot and armpit share another temporal pattern, corresponding to the blue curve in Figure~\ref{fig_skin_2}C. A similar temporal clustering of face and arm, versus foot and armpit was observed in scores of the second PC as well. 
 
\section{Discussion}
\label{sec:conc}

We have introduced a Bayesian approach to SFPCA, providing users an efficient Bayesian model selection technique like PSIS-LOO and reliable model diagnostics methods such as examining the estimated shape parameters from PSIS-LOO and utilizing the graphical posterior predictive checks. Moreover, our Bayesian modeling approach is flexible in incorporating alternative prior distributions, for example, a t-distribution to capture heavy tails in the distribution of principal component scores $\alpha_i$, which are easily implemented in \textsf{Stan}.  The examples in Section 4 demonstrate the potential of this Bayesian approach to select the optimal model and uncover meaningful biological insights after careful model implementation and diagnostics. The first limitation of our current Bayesian SFPCA method is that it can only model one temporal measurement for different subjects over time, while microbiome data are typically comprised of thousands of microbes. This drawback would restrict the microbiome applications of this method to mainly analyses of alpha diversity, changes or differences in beta diversity, or measurement of a specific microbe. But given the flexibility in Bayesian modeling, it is feasible to extend the current model to multiple outcome measures simultaneously. The second limitation of our method is that the Bayesian implementation of SFPCA is less computationally efficient than frequentist approaches. However, with the goal of building more valid and reliable models in real data applications, our flexible modeling options, model selection and diagnostics with PSIS-LOO grants advantages over other available SFPCA approaches.  Moreover, since the SFPCA model is implemented in \textsf{Stan}, a programming language with a very active user base, the \textsf{BayesTime} R package  will be able to be updated with more efficient MCMC sampling algorithms and also incorporate other groundbreaking model selection and diagnostic techniques whenever they become available. Hence, we believe that the Bayesian approach to SFPCA will enable broader applications to a wider range of longitudinal data analysis going forward. 

\section{Acknowledgements}
\label{sec:thanks}
We thank Yoshiki Vázquez-Baeza, Tomasz Kościółek, and Antonio González for suggestions and insights on microbiome data analysis, and Jeff De Reus for advice on high performance computing.

LN was partially supported by funding from the National Institute of Health grants: NIDDK 1R01DK110541-01A1 and NIA 1P01AG052352-01A1. RK was supported by NIH under grant 1DP1AT010885, NIDDK under grant 1P30DK120515, and CCFA under grant 675191. WT was supported by NIH/NIMH under grants RF1 MH120025 and R01 MH122688. 

\bibliographystyle{JASA}

\bibliography{Bibliography-BT}
\end{document}